\documentclass[runningheads]{llncs}

\usepackage[T1]{fontenc}
\usepackage[utf8]{inputenc}
\usepackage{graphicx}
\usepackage{amsmath}
\usepackage{amssymb}
\usepackage{booktabs}
\usepackage{hyperref}
\usepackage{url}
\usepackage{xcolor}
\definecolor{ccsource}{HTML}{733743}
\definecolor{ccdiscovery}{HTML}{2E5F8A}
\usepackage{tikz}
\usepackage{subfig}
\usepackage{placeins} 
\usetikzlibrary{shapes.geometric, arrows.meta, positioning, calc, backgrounds, fit}

\graphicspath{{figures/}}

\begin{document}

\title{Measuring What the Crawler Sees:\\
       Discovery Curves, Core Persistence, and Shell Dynamics
       in Longitudinal Web Crawls}
\titlerunning{Discovery Curves and Core Persistence}

\author{Michael Paris\orcidID{0000-0003-2077-6984} \and
        Hande \c{C}elikkanat\orcidID{0000-0003-2858-5867} \and
        Luca Foppiano\orcidID{0000-0002-6114-6164}}
\authorrunning{M.\ Paris et al.}
\institute{Common Crawl Foundation, Beverly Hills, USA\\
\email{\{micha,hande,luca\}@commoncrawl.org}}

\maketitle

\begin{abstract}
A longitudinal web crawl is a sequence of partial samples of an evolving URL population.
Pairwise containment between two crawls is the standard probe; under the homogeneous urn model it recovers per-round survival $\alpha$ and coverage $c$, but treats the population as uniform and consumes one pair at a time.
We extend this analysis with the \emph{discovery curve} $U(s, T)$, the cumulative URL footprint over a sliding window of $T$ crawls starting at $s$, which under the same urn model is also a closed-form function of $(\alpha, c)$.
Containment and the discovery curve are then two projections of one process: independent fits agree on $(\alpha, c)$ when the urn is homogeneous, so any disagreement is itself a measurement.
Applied to Common Crawl (2020--2025, domain granularity) and to the German Academic Web (GAW, URL granularity), the two projections disagree on both archives, and a two-component urn with a persistent core fraction $\kappa$ alongside shell parameters $(\alpha_\partial, c_\partial)$ reconciles the disagreement.
A residual on $c_\partial$ remains, signalling that the shell itself is not homogeneous; $\kappa$ is recorded as the scalar entry point to a rank-resolved generalisation, which is left to follow-up work.
\end{abstract}

\section{Introduction}
\label{sec:intro}

Common Crawl --- monthly snapshots of the open web, 2--3 billion URLs per crawl, over 90 archives since 2013 --- is the foundational corpus for most open and commercial pretraining datasets~\cite{kaplan2020scaling,penedo2023refinedweb}, including C4~\cite{raffel2020c4}, The Pile~\cite{gao2020pile}, RefinedWeb~\cite{penedo2023refinedweb}, and RedPajama~\cite{computer2023redpajama}, and underlies the web portions of GPT, Llama, Falcon, and PaLM.
The quality of every downstream corpus is bounded by which URLs Common Crawl chose to fetch and how persistently it tracked them.
Those choices are driven by a small set of static operator knobs --- host budgets, harmonic-centrality (hc-rank) thresholds, deletion cutoffs --- applied to a 200-billion-edge webgraph that is itself constantly evolving.
Their downstream effect on coverage, persistence, and discovery cost is not directly read off the data.

The per-crawl URL counts and rolling-window unique-URL statistics that Common Crawl already publishes are sufficient observables to recover the parameters of an urn model of the crawl, at the host and domain granularity, where the operator's policy levers act, and without external ground truth.
This paper derives the closed-form mapping from those public observables to the urn parameters and applies it to Common Crawl at domain granularity and to the German Academic Web (GAW) at URL granularity, two longitudinal archives produced by crawl architectures with opposing design choices.

Prior work~\cite{paris2026estimating} established the pairwise containment $g(\Delta t) = c\,\alpha^{\Delta t}$ as a closed-form recovery of $(\alpha, c)$ under the homogeneous urn.
This estimator uses one crawl pair at a time and offers no diagnostic when the homogeneous assumption fails.
The discovery curve $U(s, T)$, defined as the cumulative number of distinct URLs observed over $T$ consecutive crawls starting at $s$, uses the full sequence, and under the same urn model is also a closed-form function of $(\alpha, c)$.
Containment and the discovery curve are therefore two projections of one process: independent fits agree on $(\alpha, c)$ when the urn is homogeneous, and the magnitude of any disagreement is itself a measurement.
The pairwise fit is recurrence-weighted and reads the persistent \emph{core}; the discovery curve is fresh-discovery-weighted and reads the ephemeral \emph{shell}.
A two-component urn that introduces a persistent core fraction $\kappa$ alongside shell parameters $(\alpha_\partial, c_\partial)$ reconciles the cross-source disagreement on both archives.

The two archives differ structurally.
GAW is a Heritrix-based focused crawl seeded from approximately 150 German academic homepages, BFS-driven and is re-harvesting from a fixed URL via hops at each round.
Common Crawl is Nutch-based and recursive, with each monthly fetch list derived from the previous \texttt{CrawlDB} and ranked by harmonic centrality.
Both admit the same urn model: the two projections agree within-dataset on a single $(\kappa, \alpha, c)$ triple, while the fitted values differ across archives, reflecting the difference in the underlying URL populations.

Section~\ref{sec:related} reviews related work.
Section~\ref{sec:method} develops the urn model, the discovery formula, the two-component decomposition, and the operator-language translation.
Section~\ref{sec:results} presents empirical results on Common Crawl and GAW.
Section~\ref{sec:discussion} covers crawl-policy implications, and Section~\ref{sec:conclusion} concludes.

\section{Related Work}
\label{sec:related}

The contribution sits at the intersection of three threads.
First, \emph{urn modeling} for coverage --- the statistical machinery that descends from capture--recapture and occupancy theory and that we use to turn crawl observables into $(\alpha, c, \kappa)$.
Second, \emph{measurement of Common Crawl}: how the operator publishes what the crawler does, and how the literature has audited that output.
Third, the \emph{impact of Common Crawl}: the role it plays as the raw substrate of large-language-model pretraining, which is what makes its coverage properties matter beyond the crawler community.

\subsection{Urn modeling}
\label{sec:related_urn}

Treating the web as a finite urn from which a crawler draws repeated samples is the same problem ecologists faced with animal populations.
The two-sample base case is Petersen--Lincoln--Chapman~\cite{Petersen1896,lincoln1930calculating,Chapman1951}; its extension to a sequence of samples is Schnabel's multi-occasion census~\cite{schnabel1938fish}, and its extension to heterogeneous catchability is Chao's nonparametric lower bound~\cite{chao1987unequal}.
The underlying urn machinery goes back to Eggenberger and P{\'o}lya~\cite{eggenberger1923polya} and is summarised for the occupancy regime in Kolchin--Sevast'yanov--Chistyakov~\cite{kolchin1978random}.
Cumulative coverage --- how much of the population has been seen after $n$ samples --- is the Good--Turing question~\cite{good1953species} and, in its text-vocabulary form, Heaps' law for vocabulary growth~\cite{heaps1978information}.
The same logic was first applied to the web to bound search-engine size and overlap by Bharat and Broder~\cite{bharat1998technique} and Lawrence and Giles~\cite{Lawrence1998}.
Closest to the present setting, Paris~\cite{paris2026estimating} shows that pairwise containment between two crawls under a homogeneous urn recovers $(\alpha, c)$.
We extend that line: instead of one crawl pair we use the full sequence as a discovery curve $U(s, T)$ from the same urn, and we promote the homogeneous fit to a two-component core/shell model~\cite{chao1987unequal,vanni2025urn} when the single-population assumption fails.

\subsection{Common Crawl metrics}
\label{sec:related_cc}

Common Crawl publishes monthly archives, approximately 2--3 billion URLs each, with operator-side statistics --- per-crawl URL counts, host counts, language distribution, and rolling unique-URL footprints over the last $N$ crawls.\footnote{\url{https://commoncrawl.org/statistics}}
These descriptive metrics document the crawl as it ships; they do not themselves model what is being sampled.
Critical readings have audited what the corpus excludes by construction: Stolz and Hepp~\cite{Stolz} on structured-data pitfalls, Baack~\cite{baack2024critical} on the operational filters, and Dodge et al.~\cite{dodge2021documenting} on demographic and source skew once the data reaches downstream corpora.
Closer to longitudinal measurement, Thompson~\cite{thompson2024improved} proposes an improved methodology for treating Common Crawl as a longitudinal source --- explicitly addressing the cross-crawl comparability that any time-series analysis on top of CC must confront.
The longitudinal-measurement literature on web crawls more broadly --- per-page change-rate modelling~\cite{cho2000synchronizing,brewington2000dynamic}, freshness-aware refresh policies~\cite{Cho:2000:EWI:645926.671679,olston2010web}, identification of genuinely new pages~\cite{Toyoda:2006:WRN:1135777.1135815}, and URL persistence over multi-year windows~\cite{koehler2002web,gomes2006modelling} --- frames the crawler itself as the system to model rather than the crawl output.
Our work is closer in spirit to that thread, but inverted: we read the parameters of the underlying sampling process \emph{from} the already-published crawl observables, without instrumenting the crawler.

\subsection{Impact: Common Crawl as LLM substrate}
\label{sec:related_impact}

Kaplan et al.~\cite{kaplan2020scaling} showed that pretraining performance scales with corpus size, and Common Crawl --- the largest open web source --- became the standard substrate of open pretraining pipelines.
Direct CC-derived corpora include C4~\cite{raffel2020c4}, The Pile~\cite{gao2020pile}, OSCAR~\cite{ortizsuarez2020monolingual}, RefinedWeb~\cite{penedo2023refinedweb}, RedPajama~\cite{computer2023redpajama,weber2024redpajama}, Dolma~\cite{soldaini2024dolma}, FineWeb~\cite{penedo2024fineweb}, DCLM~\cite{li2024datacomp}, and most recently Nemotron-CC~\cite{su2024nemotroncc}; each treats Common Crawl as ground truth and competes on filtering --- language ID (\textsc{CommonLID}~\cite{suarez2026commonlid} and downstream variants), deduplication, quality classification, perplexity scoring.
The web portions of GPT, Llama, Falcon, and PaLM all sit on top of this stack.
Filter quality, however, is bounded by fetch coverage: if a URL is never fetched, no downstream filter can recover it.
The coverage of the \emph{upstream} fetch policy --- which URLs Common Crawl chose, how persistently it tracked them, what fraction of the live web that produces --- is the target measurement of the present paper, read from the crawler's own output at the (host, domain) granularity where operator policy levers act.

\section{Method}
\label{sec:method}

\subsection{Revisiting the Urn Model}
\label{sec:urn}

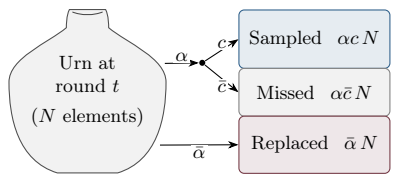
\begin{figure}[!htb]
\centering
\begin{tikzpicture}[scale=0.8, every node/.append style={transform shape},
    >={Stealth[length=1.4mm, width=1.0mm]},
    font=\small,
    box/.style={draw, rounded corners=2pt, align=center, inner sep=3pt},
    sampled/.style={box, fill=ccdiscovery!12, draw=ccdiscovery},
    missed/.style={box, fill=black!5,         draw=black!50},
    replaced/.style={box, fill=ccsource!12,   draw=ccsource},
    branch/.style={circle, fill=black, inner sep=1pt},
    elabel/.style={font=\footnotesize, inner sep=1pt, fill=white}
]
    \path[draw=gray!60!black, fill=gray!10, line join=round]
        (1.0, 0)
        -- (1.6, 0)
        .. controls (1.65, -0.10) and (1.65, -0.25) .. (1.65, -0.30)
        .. controls (1.95, -0.55) and (2.60, -0.80) .. (2.60, -1.20)
        .. controls (2.60, -1.80) and (2.45, -2.50) .. (1.80, -2.70)
        -- (0.80, -2.70)
        .. controls (0.15, -2.50) and (0.00, -1.80) .. (0.00, -1.20)
        .. controls (0.00, -0.80) and (0.65, -0.55) .. (0.95, -0.30)
        .. controls (0.95, -0.25) and (0.95, -0.10) .. (1.00, 0)
        -- cycle;
    \draw[gray!50!black, line width=0.3pt]
        (1.05, -0.04) .. controls (1.20, -0.10) and (1.40, -0.10) .. (1.55, -0.04);
    \node[align=center, font=\small] at (1.3, -1.35)
        {Urn at\\round $t$\\[3pt]{\footnotesize ($N$ elements)}};

    \node[sampled,  minimum width=2.5cm, minimum height=0.965cm, anchor=north west]
        (samp) at (3.8,  0)        {Sampled \;\,$\alpha c\,N$};
    \node[missed,   minimum width=2.5cm, minimum height=0.790cm, anchor=north west]
        (miss) at (3.8, -0.965)    {Missed \;\,$\alpha\bar c\,N$};
    \node[replaced, minimum width=2.5cm, minimum height=0.945cm, anchor=north west]
        (rep)  at (3.8, -1.755)    {Replaced \;\,$\bar\alpha\,N$};

    \node[branch] (sb) at (3.2, -0.88) {};
    \draw[->] (2.58, -0.89) -- (sb)
        node[elabel, midway, above]{$\alpha$};
    \draw[->] (sb) -- (samp.west)  node[elabel, midway, above]{$c$};
    \draw[->] (sb) -- (miss.west)  node[elabel, midway, below]{$\bar c$};
    \draw[->] (2.50, -2.24) -- (rep.west)
        node[elabel, midway, below]{$\bar\alpha$};
\end{tikzpicture}
\caption{Per-round area decomposition of the urn. The left box is the urn at round $t$, of size $N$; each element falls into one of three outcomes whose rectangle heights are proportional to the joint probabilities. \emph{Sampled} elements ($\alpha c\,N$) appear in the round-$t$ crawl; \emph{missed} elements ($\alpha\bar c\,N$) persist unseen and remain eligible for later rounds; \emph{replaced} elements ($\bar\alpha\,N$) leave the urn and are overwritten by fresh, never-seen elements. The top two bands together form the survived count $\alpha N$, and the three areas sum to $N$.}
\label{fig:urn_fates}
\end{figure}

We model a crawl as repeated sampling from a dynamic population.
An \emph{urn} contains $N$ unique elements.
Each round $t = 0, 1, 2, \ldots$ consists of two steps:
\begin{enumerate}
    \item \textbf{Sampling:} Draw a fraction $c$ of the urn uniformly (the \emph{coverage} per round).
    \item \textbf{Churning:} Retain a fraction $\alpha$ of the urn; replace the remaining $\bar\alpha \equiv 1-\alpha$ with $N\bar\alpha$ brand-new elements never seen before.
\end{enumerate}
We use the bar shorthand $\bar c \equiv 1 - c$ for the per-round miss rate and $\bar\alpha \equiv 1 - \alpha$ for the per-round churn rate throughout.
In every round, each element in the urn meets exactly one of three fates (Fig.~\ref{fig:urn_fates}): it is \emph{replaced} (churned out, probability $\bar\alpha$); or it survives and is \emph{sampled} into the crawl (probability $\alpha c$); or it survives but is \emph{missed} in the sample, persisting unseen into the next round (probability $\alpha\bar c$).
The three probabilities sum to one.

The persists-unseen rate $\alpha\bar c = \alpha(1-c)$ controls the system's memory; its complement $1 - \alpha(1-c) = \bar\alpha + \alpha c$ is the \emph{resolved rate} --- the per-round probability that an element \emph{doesn't persist unseen}, either churning out or being observed.

\subsection{Containment: Pairwise crawl intersections}
\label{sec:containment}

Prior work~\cite{paris2026estimating} showed that the pairwise containment --- the fraction of one crawl's URLs that reappear in another crawl $\Delta t$ rounds later --- has a closed form under the urn model:
\begin{equation}\label{eq:containment}
    g(\Delta t) \;=\; \frac{|u_i \cap u_j|}{|u_i|}
    \;=\; c \cdot \alpha^{\Delta t}\,,
\end{equation}
where $|u_i| = M = Nc$ is the per-crawl sample size.
A log-linear regression of $g$ against $\Delta t$ recovers the two parameters $(\alpha, c)$ from pairwise data alone.
This is the starting point; the question is what happens when we move from pairs to \emph{sequences}.

\subsection{Discovery Curve: Growth of unique elements}
\label{sec:discovery_curve}

Let $\rho(t)$ denote the fraction of the urn that has been seen in at least one of the rounds $0, \ldots, t-1$, with $\rho(0) = 0$.
Two streams of revisits enter round $t$, both filtered by the per-element survival factor $\alpha$ --- prior revisits $\rho(t-1)$ and the prior round's new samples $c\,\nu(t-1)$ that just got marked seen:
\begin{equation}\label{eq:recurrence}
    \rho(t) \;=\; \alpha\bigl(\rho(t-1) + c\,\nu(t-1)\bigr)
            \;=\; \alpha c \;+\; \alpha\bar c\,\rho(t-1)\,.
\end{equation}
This is a linear recurrence with driving term $\alpha c$ and ratio $\alpha\bar c$ (the persists-unseen rate).
Using the identity $\bar\alpha + \alpha c = 1 - \alpha\bar c$, its solution is
\begin{equation}\label{eq:rho_closed}
    \rho(t) \;=\; \frac{\alpha c\,\bigl(1 - (\alpha\bar c)^t\bigr)}{1 - \alpha\bar c}\,.
\end{equation}

The \emph{discovery curve} is the cumulative sum of distinct elements observed through round $t$:
\begin{equation}
    \hat{U}(t) \;=\; \sum_{\tau=0}^{t} \Delta\hat{U}(\tau)\,,
\end{equation}
where the per-round new fraction $\nu(t) \equiv \Delta\hat{U}(t)/M = 1 - \rho(t)$ counts new discoveries in round $t$.
Summing the geometric series gives the closed form
\begin{equation}\label{eq:Uclosed}
	\boxed{\;\frac{\hat{U}(t)}{M}
		\;=\; \frac{(t+1)\,\bar\alpha}{\bar\alpha + \alpha c}
		\;+\; \frac{\alpha c\,\bigl(1 - (\alpha\bar c)^{t+1}\bigr)}{(\bar\alpha + \alpha c)^2}\;}
\end{equation}
This is the \emph{discovery formula}: a constant intercept, a linear trend with slope $\nu_\infty = \bar\alpha/(1-\alpha\bar c)$ (the asymptotic new-discovery rate), and a geometric transient that decays with ratio $\alpha\bar c$ and vanishes as $t\to\infty$.
The persists-unseen rate $\alpha\bar c$ is the only memory parameter that appears: as the transient's ratio, and via its complement $1-\alpha\bar c$ in every denominator.
The same $(\alpha, c)$ that govern the pairwise containment~\eqref{eq:containment} now predict the entire growth curve of the discovery curve --- a substantially richer observable derived from the same two parameters.

\paragraph{Discrete derivative and $\kappa$-identifiability.}
\label{sec:newly}
The per-round increment $\Delta\hat{U}(t)/M$ is the discrete derivative of $\hat{U}(t)/M$ and carries no information independent of the discovery curve.
Fitting the two-component model against the increment alone collapses to the homogeneous case ($\kappa \to 0$ for $t \geq 1$), because the core contributes no new mass after the first crawl.
We use the increment as a consistency check only: any well-resolved $\kappa$ must be anchored from the containment floor (\S\ref{sec:two_component}), not from the accumulation slope.
The cross-source diagnostic in \S\ref{sec:results} therefore reports fits on (containment, discovery curve) only.

\subsection{Asymptotic Regime}
\label{sec:asymptote}

For large $t$, $(\alpha\bar c)^{t+1} \to 0$ and the progressive union becomes linear:
\begin{equation}\label{eq:asymptote}
    \frac{\hat{U}(t)}{M} \;\approx\; \nu_\infty\,t \;+\; I\,,
\end{equation}
with slope $\nu_\infty = \bar\alpha/(1-\alpha\bar c)$ and intercept
\begin{equation}\label{eq:I_def}
    I \;=\; 1 \;+\; \rho_\infty\,\tau_h\,,
    \qquad \tau_h \;\equiv\; \frac{\alpha\bar c}{1 -\alpha\bar c}\,,
\end{equation}
where $\tau_h$ is the \emph{hiding time per resolution} (the average length of a single hidden epoch).
The intercept $I$ is the asymptotic line \emph{extrapolated back} to $t = 0$ --- $I = \hat U_{\rm asy}(0)/M$ where $\hat U_{\rm asy}(t) \equiv (\nu_\infty t + I)\,M$.
The actual cumulative ratio at $t = 0$ is $\hat U(0)/M = 1$ (the first crawl is all-new); the difference $I - 1 = \rho_\infty\,\tau_h$ is the integrated transient --- revisit rate times hiding time.

Inverting from the observable pair $(\nu_\infty, \tau_h)$ to model parameters $(\alpha, c)$, with $\rho_\infty = 1 - \nu_\infty$:
\begin{align}
    \alpha &\;=\; \frac{\rho_\infty + \tau_h}{1 + \tau_h}\,,
        \label{eq:ainv}\\
    c      &\;=\; \frac{\rho_\infty}{\rho_\infty + \tau_h}\,.
        \label{eq:cinv}
\end{align}
The discovery curve thus admits \emph{two} independent homogeneous fits from its own data: an analytic regression of the full discovery formula~\eqref{eq:Uclosed}, and an asymptotic inversion of the linear tail.
When the two agree, the population is well-described by a single $(\alpha, c)$; when they diverge, the transient and the tail are sampling different sub-populations --- the first sign that a two-component model is needed.

\subsection{Operator language: lifetime, sampled time, hidden time}
\label{sec:operator_translation}

The urn parameters $(\alpha, c)$ are abstract per-round or per-time unit probabilities.
The operator measures crawls.
Three derived quantities translate $(\alpha, c)$ into operator-facing units of crawls per URL life:
\begin{align}
    \ell      &\;\equiv\; \frac{\alpha}{\bar\alpha},
        \label{eq:life_expectancy}\\
    \eta      &\;\equiv\; c\,\ell \;=\; \frac{\alpha c}{\bar\alpha},
        \label{eq:sampled_time}\\
    \bar\eta  &\;\equiv\; \bar c\,\ell \;=\; \frac{\alpha\bar c}{\bar\alpha},
        \label{eq:hiding_time}
\end{align}
respectively the \emph{life expectancy} (expected crawls a URL persists in the urn), \emph{sampled time} (expected crawls a URL is re-fetched over its life), and \emph{hidden time} (expected crawls a URL is alive but unsampled).
By construction, the lifetime decomposes additively into time spent sampled and time spent hidden, and the sampling-to-hiding ratio reduces to the per-round coverage odds:
\begin{align}
    \ell &\;=\; \eta + \bar\eta\,,
        \label{eq:lifetime_decomposition}\\
    \frac{\eta}{\bar\eta} &\;=\; \frac{c}{\bar c}\,.
        \label{eq:sample_hide_ratio}
\end{align}
A complementary timescale, the \emph{hiding time per resolution} $\tau_h$ from~\eqref{eq:I_def}, measures the average length of a single hidden epoch (rounds a hidden URL stays hidden before being sampled or churned out).
The two are related by
\begin{equation}\label{eq:hide_per_life_per_epoch}
    \bar\eta \;=\; (1 + \eta)\,\tau_h\,,
\end{equation}
because a URL's total hidden time equals the per-epoch hiding time times the number of hidden epochs --- one before each sample plus a final one ending in churn.
$\tau_h$ governs the asymptotic intercept~\eqref{eq:I_def}; $\bar\eta$ measures the per-URL hidden cost over the entire life.

This is the bridge between the model and the operator.
The web supplies the lifetime $\ell$ (URL persistence in the operator's urn); the policy levers --- host budgets, hc-score thresholds, recrawl rules, deletion behaviour --- control how that lifetime splits between sampled rounds and hidden rounds via $c$.
The cross-source diagnostic loop introduced in \S\ref{sec:intro} returns its readings in $(\alpha, c)$; equations \eqref{eq:life_expectancy}--\eqref{eq:sample_hide_ratio} convert those readings into the language an operator uses to plan a crawl.

\subsection{Two-Component Decomposition: Core $K$ and Shell $\partial K$}
\label{sec:two_component}

When the homogeneous fits from the two projections --- \eqref{eq:containment} on containment and \eqref{eq:Uclosed} on the discovery curve --- disagree on $(\alpha, c)$, the disagreement is the signature of a heterogeneous urn.
The minimal extension that closes the gap is a two-component partition: a persistent \emph{core} $K$ of size $\kappa N$ with $(\alpha_K, c_K) = (1, 1)$ --- immortal and fully sampled --- and an ephemeral \emph{shell} $\partial K$ of size $(1 - \kappa) N$ with parameters $(\alpha_\partial, c_\partial)$.
The core is graph-driven (the well-linked URLs that any reasonable policy keeps visible); the shell carries the policy-shaped ephemeral mass.

\paragraph{Containment.}
The core contributes a persistent floor at $\kappa$ to pairwise containment:
\begin{equation}\label{eq:g_two_component}
    \boxed{\;g(\Delta t) \;=\; \kappa \;+\; (1-\kappa)\,c_\partial\,\alpha_\partial^{\Delta t}\;}
\end{equation}
On a log scale this curve is concave; a single-exponential fit cannot capture it, and a joint 3-parameter regression of \eqref{eq:g_two_component} recovers $(\kappa, \alpha_\partial, c_\partial)$.

\paragraph{Per-crawl core share.}
Each crawl draws $N\kappa$ core URLs (always sampled) and $N(1-\kappa)c_\partial$ shell URLs, giving a crawl size $M = N\kappa + N(1-\kappa)\,c_\partial$.
The fraction of each crawl composed of core mass is
\begin{equation}\label{eq:mu_def}
    \mu \;\equiv\; \frac{N\kappa}{M}
    \;=\; \frac{\kappa}{\kappa + (1-\kappa)\,c_\partial}\,.
\end{equation}
Because $c_\partial < 1$ while $c_K = 1$, $\mu > \kappa$ --- the core is \emph{over-represented} in any single crawl relative to its urn share.

\paragraph{Discovery curve.}
The cumulative-unique count splits additively.
The core saturates with $N\kappa$ URLs at $t = 0$ and contributes nothing thereafter; the shell follows the homogeneous discovery formula~\eqref{eq:Uclosed} with its own parameters.
Letting $B(t;\,\alpha,c)$ denote the right-hand side of~\eqref{eq:Uclosed}:
\begin{equation}\label{eq:U_two_component}
    \boxed{\;\frac{\hat{U}_{2c}(t)}{M} \;=\; \mu \;+\; (1-\mu)\,B\bigl(t;\,\alpha_\partial, c_\partial\bigr)\;}
\end{equation}
The per-round new fraction is $(1-\mu)\,\Delta B(t; \alpha_\partial, c_\partial)$ for $t > 0$ --- every newly-discovered URL after the first crawl is shell mass.

\paragraph{Identifiability and the two triples.}
$\kappa$ is most cleanly anchored from the containment floor in \eqref{eq:g_two_component}, where the core contributes statically at every $\Delta t$.
The discovery curve admits a joint 3-parameter regression on \eqref{eq:U_two_component}, but the core plateau and the shell discovery formula trade off along a shallow ridge --- so the discovery-curve $\kappa$ is the looser of the two anchors.
Figs.~\ref{fig:cc_two_component} and~\ref{fig:gaw_two_component} report the two triples on Common Crawl and the German Academic Web respectively; agreement on $\kappa$ across the two projections is the diagnostic, and the residual disagreement on $c_\partial$ at short timescales is read as evidence of a structurally non-uniform shell that a single scalar cannot capture.

\paragraph{Operator language at the shell.}
The lifetime decomposition of \S\ref{sec:operator_translation} applies componentwise.
The core has $\ell_K \to \infty$ and is fully resolved each round ($\eta_K = \ell_K$, $\bar\eta_K = 0$); the shell carries finite $\ell_\partial = \alpha_\partial/(1 - \alpha_\partial)$, $\eta_\partial = c_\partial\,\ell_\partial$, $\bar\eta_\partial = (1 - c_\partial)\,\ell_\partial$.
Policy levers act primarily on the shell triple while leaving $\kappa$ relatively stable.

\section{Results}
\label{sec:results}

\subsection{Common Crawl: The Discovery-Curve Object}
\label{sec:results_cc_discovery_curve}

\begin{figure*}[!htb]
\centering
\subfloat[Heatmap $U(s, T)$ (lag $T$ vs.\ crawl date)\label{fig:cc_heatmap_panel}]{%
    \includegraphics[width=0.48\textwidth]{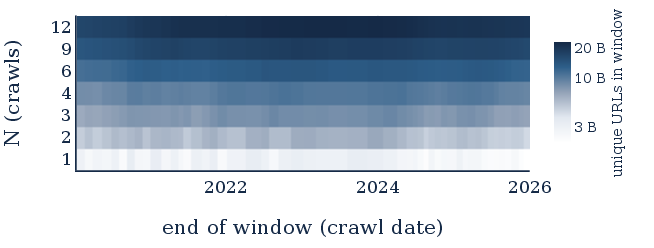}%
}\hfill
\subfloat[Normalised $U(\Delta t)/U(0)$ vs.\ window length\label{fig:cc_per_start_norm_panel}]{%
    \includegraphics[width=0.48\textwidth]{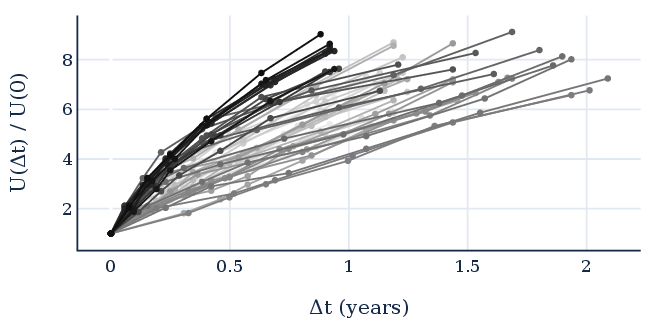}%
}
\caption{Discovery curve on Common Crawl, 2020--2025.
    \textbf{(a)}~Heatmap $U(s, T)$ over the monthly archive: lag $T$ on the $y$-axis, starting crawl date $s$ on the $x$-axis, cell colour = unique URLs in the rolling window $[s, s+T]$.
    The public \texttt{url\_last\_N} statistic is the row at fixed $T$.
    \textbf{(b)}~Normalised $U(\Delta t)/U(0)$ against the actual elapsed time $\Delta t$ in years (rather than the crawl count $N$, since CC cadence ranges from biweekly to occasional 1--2 month gaps in 2021--2022, so a $N=12$ window can span anywhere from 0.9 to 2.1 years), one line per ending crawl (grey = older, black = newer): dividing by the single-crawl cardinality collapses the cross-start differences and isolates the discovery-formula shape under the urn model --- the regression target for Figs.~\ref{fig:cc_homogeneous}--\ref{fig:cc_two_component}.
    The 2020--2025 window is the same interval used in every subsequent CC panel.}
\label{fig:cc_discovery_curve}
\end{figure*}

Common Crawl publishes a per-crawl URL count and a family of rolling-window unique-URL counts \texttt{url\_last\_N} for $N \in \{1, 2, 3, 4, 6, 9, 12\}$ on the operator's crawl-statistics page.\footnote{\url{https://commoncrawl.github.io/cc-crawl-statistics/plots/crawlsize}} Pivoted across all valid (start, window-length) pairs in the 2020--2025 monthly archive, these counts trace the discovery curve $U(s, T)$ as a function of the starting crawl $s$ and window length $T$ (Fig.~\ref{fig:cc_discovery_curve}a).
Two features come straight off the figure.
First, $U$ grows monotonically in $T$ but with a clearly decelerating slope --- the hallmark of a finite urn with revisits.
Second, dividing each curve by its single-crawl cardinality collapses the family across starts (Fig.~\ref{fig:cc_discovery_curve}b), confirming that on this window the per-start trajectories are well-described by a common normalised shape.
That common shape is the regression target for the cross-source fits below.

The normalised discovery curve is a single projection of the underlying sampling process; pairwise containment $g(\Delta t)$ is the second.
Under a homogeneous urn both are governed by the same $(\alpha, c)$, so fitting each independently and overlaying both fits on both observables turns cross-source agreement into a falsifiability test --- where the two projections disagree, the homogeneous urn is the wrong model, and the size and shape of the gap dictate the minimal extension.
The remainder of this section runs that test at \emph{domain} granularity, the unit at which CC's host-budget and harmonic-centrality levers actually act, and the result shapes \S\ref{sec:results_gaw} (closed-archive validation).

\subsection{Common Crawl: Cross-Source Fits at Domain Granularity}
\label{sec:results_cc_cross_source}

Both projections are reduced to the same 2020--2025 monthly window used in Fig.~\ref{fig:cc_discovery_curve} and aggregated to the domain population.
Each panel below uses the parameters fit on \emph{one} source --- containment in red, discovery curve in blue --- but overlays them on \emph{both} observables, so the cross-source gap is read off directly: cross-source agreement looks like two overlapping curves on each panel, disagreement is the gap between them.

\begin{figure}[htbp]
\centering
\subfloat[$U^{(d)}(T)/U^{(d)}(0)$ vs.\ $T$\label{fig:cc_m1_pu}]{%
    \includegraphics[width=0.47\columnwidth]{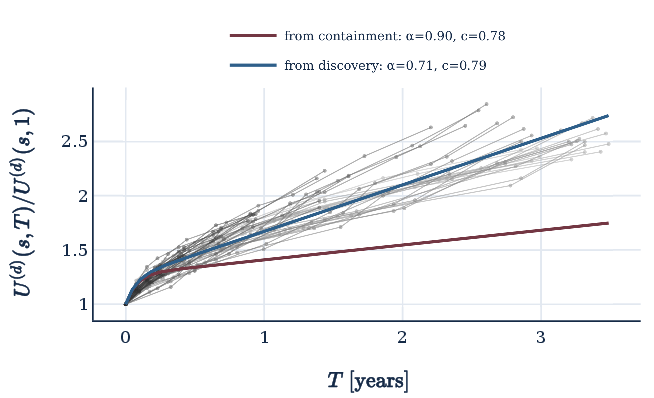}%
}\hfill
\subfloat[Containment $g^{(d)}(\Delta t)$ vs.\ $\Delta t$\label{fig:cc_m1_cont}]{%
    \includegraphics[width=0.47\columnwidth]{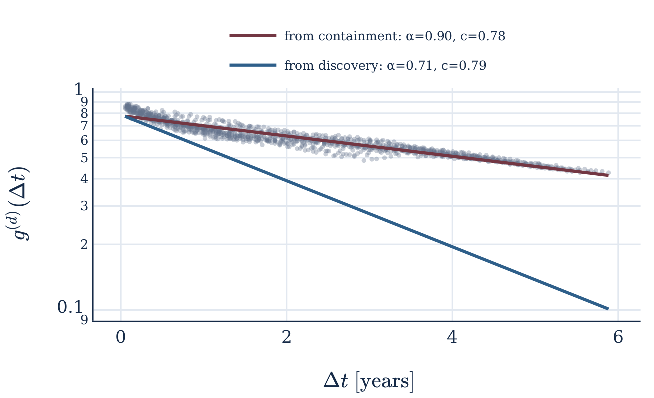}%
}
\caption{\textbf{Common Crawl --- homogeneous cross-source view at domain granularity, 2020--2025.} Both panels overlay the domain-level data with two homogeneous urn curves: \textcolor{ccsource}{burgundy} uses $(\alpha_g, c_g)$ fit on containment, \textcolor{ccdiscovery}{navy} uses $(\alpha_U, c_U)$ fit on the discovery curve.
    If the domain population were homogeneous, the two curves would coincide on both panels; the size of the gap is the homogeneous-fit failure on this window and motivates the two-component extension in Fig.~\ref{fig:cc_two_component}.}
\label{fig:cc_homogeneous}
\end{figure}

\begin{figure}[htbp]
\centering
\subfloat[$U^{(d)}(T)/U^{(d)}(0)$ vs.\ $T$\label{fig:cc_m2_pu}]{%
    \includegraphics[width=0.47\columnwidth]{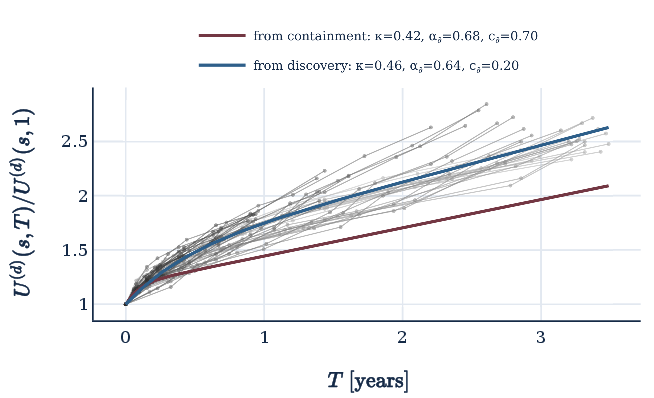}%
}\hfill
\subfloat[Containment $g^{(d)}(\Delta t)$ vs.\ $\Delta t$\label{fig:cc_m2_cont}]{%
    \includegraphics[width=0.47\columnwidth]{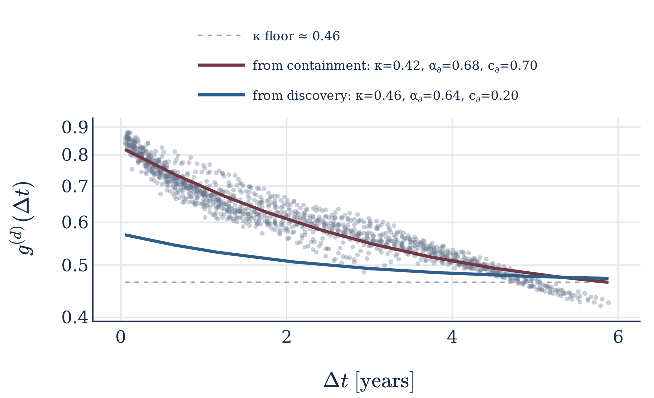}%
}
\caption{\textbf{Common Crawl --- two-component cross-source view at domain granularity, 2020--2025.} Same domain observables as Fig.~\ref{fig:cc_homogeneous}, overlaid now with two-component fits: \textcolor{ccsource}{$(\alpha_{\partial,g}^{(d)}, c_{\partial,g}^{(d)}, \kappa_g^{(d)})$} from containment and \textcolor{ccdiscovery}{$(\alpha_{\partial,U}^{(d)}, c_{\partial,U}^{(d)}, \kappa_U^{(d)})$} from the discovery curve.
    The two fits agree on a persistent core fraction $\kappa^{(d)} \approx 0.4$ --- cross-source convergence on the mass split that the homogeneous urn cannot resolve.
    The residual disagreement on $c_\partial^{(d)}$ reflects the asymmetry between containment (recurrence-weighted) and the discovery curve (fresh-discovery-weighted) --- a signature of a structurally non-uniform shell that a single scalar cannot capture.}
\label{fig:cc_two_component}
\end{figure}

\begin{figure}[htbp]
\centering
\subfloat[$U(T)/U(0)$ vs.\ $T$\label{fig:gaw_m2_pu}]{%
    \includegraphics[width=0.49\columnwidth]{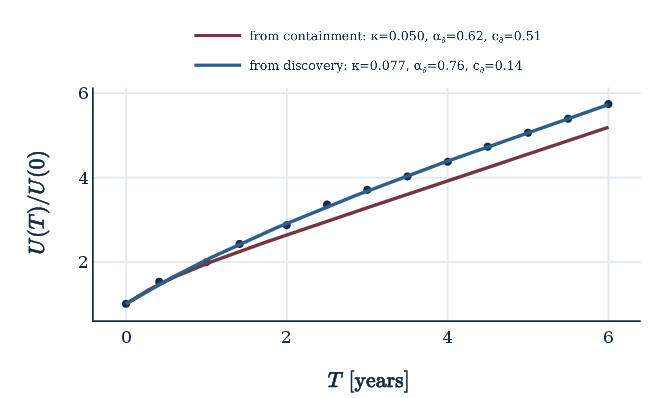}%
}\hfill
\subfloat[Containment $g(\Delta t)$ vs.\ $\Delta t$\label{fig:gaw_m2_cont}]{%
    \includegraphics[width=0.49\columnwidth]{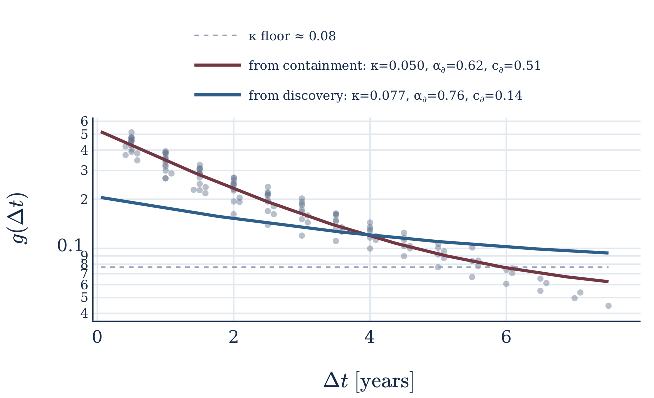}%
}
\caption{\textbf{German Academic Web --- two-component validation.} The same two-component cross-source check as Fig.~\ref{fig:cc_two_component}, applied to the closed-archive GAW.
    GAW supplies a single longitudinal trajectory rather than a family indexed by window-start (the closed archive has one fixed start), so this is a single-trajectory cross-source test; cross-source agreement on a single $(\kappa, \alpha, c)$ triple is the ground-truth check on the framework before transferring it to Common Crawl.
    Fitted $(\kappa, \alpha, c)$ values are annotated in the figure legends.}
\label{fig:gaw_two_component}
\end{figure}

The two homogeneous fits diverge visibly on both panels of Fig.~\ref{fig:cc_homogeneous}: containment pulls $\alpha$ high and $c$ moderate (it is dominated by the persistent recurring mass), the discovery curve pulls them lower (it is dominated by churn-driven fresh discoveries).
Neither single $(\alpha, c)$ explains both projections at once.
The minimal extension that closes this gap is the two-component urn of \S\ref{sec:two_component}: a persistent core of fraction $\kappa$ (immortal, fully sampled) plus a shell with its own $(\alpha_\partial, c_\partial)$.
Refitting both projections under this model gives Fig.~\ref{fig:cc_two_component}.

The two-component fits on Fig.~\ref{fig:cc_two_component} land on a common $\kappa^{(d)} \approx 0.4$ from both projections --- the mass split that the homogeneous urn could not see.
The shell parameters $(\alpha_\partial^{(d)}, c_\partial^{(d)})$ also tighten substantially, though a residual containment-vs-discovery gap on $c_\partial^{(d)}$ remains; that gap is the empirical entry point to a rank-resolved treatment of the shell, which we return to briefly in \S\ref{sec:discussion}.
Before doing so we ask whether the same two-projection procedure --- and the same structural prediction (cross-source convergence on a single $(\kappa, \alpha, c)$ triple) --- transfers to a crawler built on opposing design philosophies.

\subsection{Validation on the German Academic Web}
\label{sec:results_gaw}

The German Academic Web (GAW) is a closed Heritrix-based focused crawl seeded from German academic homepages, biannual rather than monthly, with URL-level snapshots over 2013--2019.
Its design philosophy is the opposite of CC's: a fixed seed hopper, breadth-first traversal, no harmonic-centrality re-ranking.
Repeating the two-projection two-component check on this archive therefore tests whether the structural prediction --- cross-source convergence on a single $(\kappa, \alpha, c)$ triple --- is a property of the framework or an artefact of the CC pipeline.
GAW supplies a single longitudinal trajectory rather than a family indexed by window-start $s$ (the closed archive has one fixed start), so the cross-source test here is single-trajectory; its value is in whether the two projections agree at all, not in the per-trajectory variance.

The fitted $(\kappa, \alpha, c)$ on GAW differ numerically from the CC-domain values --- the two crawls sample different web populations and operate at different cadences --- but the structural signature holds: both projections converge on the same triple, and the homogeneous reading of either projection alone is biased in the same direction as on Common Crawl.
The framework transfers across architectures.
What it does not yet resolve is the residual disagreement on $c_\partial$ visible in Fig.~\ref{fig:cc_two_component}, nor whether the persistent core is a single rigid block or has internal structure of its own.
Both questions live on the rank axis, and we sketch the path to a rank-resolved $\kappa(r)$ profile in the discussion, leaving the full development to follow-up work.

\FloatBarrier

\section{Discussion}
\label{sec:discussion}

The question this paper started with was whether the operator can infer the effective sampling fraction $c$, persistence $\alpha$, and core/shell decomposition $\kappa$ from the crawler's own output, at the (host, domain) granularity where policy levers actually act, and without external ground truth.
Two cross-source diagnostics on the 51-crawl 2020--2025 Common-Crawl-domain window answer yes: homogeneous fits to containment and to the discovery curve disagree visibly on $(\alpha, c)$; the two-component fit reconciles them on a common $\kappa^{(d)} \approx 0.4$ and reduces the gap on $(\alpha_\partial^{(d)}, c_\partial^{(d)})$ to a single residual on $c_\partial$.
The same procedure on the closed-archive German Academic Web converges on its own triple $(\kappa, \alpha, c) \approx (0.06, 0.62, 0.51)$ at URL level, and the cross-source structural prediction --- two projections, one triple --- transfers across crawl architectures.

\paragraph{What the urn parameters represent.}
The asymptotic regime of the discovery formula gives the operator two regression-only observables --- the asymptotic new-discovery slope $\nu_\infty$ and the long-run seen-fraction $\rho_\infty = 1 - \nu_\infty$ --- and inversion sends $(\nu_\infty, \rho_\infty)$ to $(\alpha, c)$.
Read this way $c$ is the per-round coverage fraction --- directly tunable by the operator through host budgets, hc-rank thresholds, and revisit cadence; $\alpha$ is the per-round survival rate of the urn, which the operator does not pick (the web supplies it through the persistence of the eligible URL set); and $\kappa$ is the joint readout of which URLs the policy keeps eligible (operator) and how stable that eligible set is on the web (web).
Empirically the conjecture lines up: $c$ moves visibly between crawlers (CC-domain $\approx 0.78$, GAW-URL $\approx 0.51$) tracking the operating-point difference, while $\alpha$ moves with what the population actually retains; calling out this split is a conjecture, not a theorem, and tying it down would require holding one side fixed while varying the other.
A subtlety on $\alpha$: the urn has no rebirth mechanism, so any apparent comeback in the data --- a URL absent for several crawls and reappearing later --- folds in either as an extended run of survived-and-missed events (long $\ell$, high $\bar\eta$) or as a fresh element happening to share the identifier; the two are observationally indistinguishable and both fold via the same $\alpha^{\Delta t}$ kernel.
The recovered $\alpha$ is therefore not a strict web survival rate but the survival rate the urn sees --- a crawl-side readout that absorbs short dormancies into the hidden-time term and only registers operator-visible disappearances as churn.

\paragraph{Inhomogeneity, $\kappa$, and a rank-resolved outlook.}
The scalar $\kappa$ recovered here is the simplest possible summary of population heterogeneity: a mass split between a fully resolved core and a homogeneously churning shell.
The residual containment-vs-discovery gap on $c_\partial^{(d)}$ visible in Fig.~\ref{fig:cc_two_component} is the signal that the shell itself is not homogeneous --- containment, being recurrence-weighted, anchors on the inner shell, while the discovery curve, being fresh-discovery-weighted, drags toward the outer shell.
The natural extra observable is the per-domain rank change between consecutive crawls on the operator's hc-rank axis, which promotes the scalar $\kappa$ to a continuous rigidity profile $(\mu(r), D(r))$ --- the same axis on which host budgets and harmonic-centrality thresholds act.
The full development --- drift, diffusion, and a survivor-flux balance that ties together the four rank-resolved profiles --- is left to follow-up work; here we record only that the framework's diagnostic chain (discovery curve $\to$ cross-source disagreement $\to$ $\kappa$) terminates in a scalar that has a natural lift to the operator's policy axis, and that the residual on $c_\partial$ is its first measurable consequence.

\section{Conclusion}
\label{sec:conclusion}

This paper extends the language an operator has for talking about a crawl.
From the same per-crawl URL counts that operators already publish, two projections of the urn process --- pairwise containment and the discovery curve --- pin down a coverage fraction $c$ (what share of the urn each round samples), a churn fraction $\bar\alpha = 1 - \alpha$ (what share of the urn each round replaces), an asymptotic unique rate $\nu_\infty = \bar\alpha/(\bar\alpha + \alpha c)$ (what fraction of each new crawl is genuinely new), and a core fraction $\kappa$ (what share of the urn the policy keeps fully resolved).

Recombined, the same parameters give the per-URL life expectancy $\ell = \alpha/\bar\alpha$ in crawls, the resolve time $\eta = c\,\ell$ (crawls a URL is re-fetched over its life), and the hidden time $\bar\eta = \bar c\,\ell$ (crawls a URL is alive but unsampled).
A crawl that was previously a sequence of raw counts becomes a small dictionary of comparable quantities: coverage, churn, unique rate, life expectancy, resolve time, hidden time, core fraction.
The dictionary is the contribution.

\begin{credits}
\subsubsection{\discintname}
The authors have no competing interests to declare.
\end{credits}


\FloatBarrier 
\bibliographystyle{splncs04}
\bibliography{bibliography}

\end{document}